\documentclass[prl,twocolumn,showpacs,preprintnumbers,amsmath,amssymb,superscriptaddress,unsortedaddress]{revtex4}

\pdfoutput=1
\usepackage{graphicx}
\usepackage{dcolumn}
\usepackage{bm}

\begin{document}

\title{Entanglement purification protocol for a mixture of a pure entangled state and a pure product state}



\author{Miko{\l}aj Czechlewski}

\affiliation{Faculty of Physics,
 Adam Mickiewicz University,
Umultowska 85, 61-614 Pozna\'{n}, Poland}

\author{Andrzej Grudka}

\affiliation{Faculty of Physics, Adam Mickiewicz University, Umultowska 85, 61-614 Pozna\'{n}, Poland}

\affiliation{Institute of Theoretical Physics and Astrophysics,
University of Gda\'{n}sk, Wita Stwosza 57, 80-952 Gda\'{n}sk, Poland}

\affiliation{National Quantum Information Centre of Gda\'{n}sk, W{\l}adys{\l}awa Andersa 27, 81-824 Sopot, Poland}

\author{Satoshi Ishizaka}
\affiliation{Nano Electronics Research Laboratories, NEC
Corporation, 34 Miyukigaoka, Tsukuba 305-8501, Japan}
\affiliation{INQIE, The University of Tokyo, 4-6-1 Komaba,
Meguro-ku, Tokyo 153-8505, Japan}

\author{Antoni W\'ojcik}

\affiliation{Faculty of Physics,
 Adam Mickiewicz University,
Umultowska 85, 61-614 Pozna\'{n}, Poland}


\date{\today}

\begin{abstract}
We present an entanglement purification protocol for a mixture of a pure entangled state and a pure product state, which are orthogonal to each other. The protocol is combination of bisection method and one-way hashing protocol.  We give recursive formula for the rate of the protocol for different states, i.e. the number of maximally entangled two-qubit pairs obtained with the protocol per a single copy of the initial state. We also calculate numerically the rate for some states.

\end{abstract}
\maketitle

\maketitle

Entanglement is resource in quantum information. Maximally entangled states are basic ingredient of fundamental quantum information protocols like e.g. quantum teleportation \cite{Bennett5}, dense coding \cite{Bennett6} or Ekert's quantum cryptographic protocol \cite{Ekert1}. In these protocols the maximally entangled pair is shared by two parties and used to perform a certain task. However, in the real world the parties share noisy entangled pairs. Bennett {\it et al.} have shown that many pairs in mixed entangled states can be distilled to a smaller number of pairs in nearly maximally entangled states \cite{Bennett1, Bennett3}. In particular, they presented purification protocols that can be realized by means of local operations and classical communication (LOCC). Let us suppose that the parties share $n$ pairs of qubits, each of which is in the state  $\rho$. If using a particular protocol the parties can obtain $m$ pairs of qubits, each of which is in the maximal entangled state, from $n$ pairs of qubits, each of which is in state $\rho$ then the protocol has the  rate $\frac{m}{n}$ for the state $\rho$. Different protocols have different rates and moreover one needs different protocols for different states. The maximal rate for state $\rho$, i.e., the rate of the optimal protocol for state $\rho$, is called distillable entanglement. Distillable entanglement of mixed states is usually difficult to calculate and it is only known for  bound entangled states, maximally correlated states and some other specific mixed states \cite{Horodecki7, Rains1, Eisert1, Chen1, Hamieh1, Hiroshima1}. For bound entangled states it is equal to zero \cite{Horodecki7}, while for maximally correlated states it is equal to the relative entropy of entanglement \cite{Vedral1, Vedral2, Rains2, Hiroshima1} and can be distilled by the one-way hashing protocol \cite{Devetak1, Devetak2}.
However, for many states only upper and lower bounds on distillable entanglement are known. Lower bounds are usually given by the rates of particular protocols. Upper bounds are given by entanglement measures known to be greater or equal to distillable entanglement, e.g. the relative entropy of entanglement. 

In this paper we present an entanglement purification protocol for a mixture of a pure entangled state and a pure product state which are orthogonal to each other. The very first protocol for these states was presented in \cite{Bennett3}. Our protocol is a combination of Procrustean method of entanglement concentration \cite{Bennett2}, bisection method and one-way hashing protocol \cite{Devetak1, Devetak2}. 

We assume that Alice and Bob share many copies of the state
\begin{equation}
\rho=p|\psi^{+}\rangle\langle\psi^{+}|+(1-p)|00\rangle\langle 00|,
\label{eq:6}
\end{equation}
where 
\begin{equation}
|\psi^{+}\rangle=\frac{1}{\sqrt{2}}(|10\rangle+|01\rangle)
\end{equation}
is the maximal entangled state. State $\rho$ is a mixture of the maximal entangled state and a product state. The product state is orthogonal to the maximal entangled state. Let Alice and Bob group these states in blocks of $n$ copies, where $n$ is the power of $2$. It is convenient to write $\rho^{\otimes n}$ in the following way
\begin{eqnarray}
& \rho^{\otimes n}=p^{n}|\psi^{+}\rangle\langle\psi^{+}|^{\otimes n}+\nonumber \\
& p^{n-1}(1-p)[|\psi^{+}\rangle\langle\psi^{+}|^{\otimes(n-1)} |00\rangle\langle 00|+\dots]+\nonumber \\
& +p^{n-2}(1-p)^{2}[|\psi^{+}\rangle\langle\psi^{+}|^{\otimes(n-2)}|00\rangle\langle 00|^{\otimes 2}+\dots]\nonumber\\
& \dots+(1-p)^{n}|00\rangle\langle 00|^{\otimes n},
\end{eqnarray}
where "$\dots$" in each square bracket stands for all permutations of the first term in the square bracket.

Let each party project her/his part of the state on a subspace spanned by vectors with the definite number of  $1$'s and definite number of $0$'s, i.e., Alice and Bob perform von Neumann measurements given by the sets of projectors
\begin{equation}
\{ P_{a}^{n}=\sum\limits_{[x]=n,|x|=a}|x\rangle\langle x| \}
\end{equation}
and
\begin{equation}
\{ P_{b}^{n}=\sum\limits_{[x]=n,|x|=b}|x\rangle\langle x| \},
\end{equation}
respectively, where $|x|$ denotes the Hamming weight of the string $x$ of $[x]=n$ qubits and $a, b \in \{0,1,...n\}$.
If Alice obtains $P_{a}^{n}$ as a result of her measurement and Bob obtains $P_{b}^{n}$ as a result of his measurement, then all terms in the expansion of $\rho^{\otimes n}$ except the term with $a+b$ $|\psi^{+}\rangle$'s and $n-a-b$ $|00\rangle$'s are annihilated, i.e., the post measurement state is 
\begin{eqnarray}
& \rho(n,a,b)= \nonumber \\
& =\frac{1}{p(n,a,b)} P_{a}^{n}P_{b}^{n}[|\psi^{+}\rangle\langle\psi^{+}|^{\otimes(a+b)}|00\rangle\langle 00|^{\otimes n-a-b}+ \nonumber\\
& +\dots]P_{a}^{n}P_{b}^{n},
\label{eq:1}
\end{eqnarray}
where
\begin{equation}
p(n,a,b)={n \choose a+b}{a+b \choose a}2^{-(a+b)}.
\label{eq:3}
\end{equation}
The probability of this event is
\begin{equation}
P(n,a,b)=p^{a+b}(1-p)^{n-a-b}p(n,a,b).
\end{equation}
The factor $p^{a+b}(1-p)^{n-a-b}{n \choose a+b}$ is the probability that Alice and Bob share the state $|\psi^{+}\rangle\langle\psi^{+}|^{\otimes(a+b)}|00\rangle\langle 00|^{\otimes n-a-b}$ or any permutation of it and ${a+b \choose a}2^{-(a+b)}=\text{Tr}[P_{a}^{n}P_{b}^{n}|\psi^{+}\rangle\langle\psi^{+}|^{\otimes(a+b)}|00\rangle\langle 00|^{\otimes n-a-b}P_{a}^{n}P_{b}^{n}]$.

If $a+b=n$, then Alice and Bob share the maximal entangled state of the rank $r(n,a)$, where
\begin{eqnarray}
r(n,a)={n \choose a},
\end{eqnarray}
i.e., they share $\log_2{r(n,a)}$ maximally entangled pairs of qubits.
If one of the equalities: $a=0$, $a=n$, $b=0$ or $b=n$ holds, then Alice and Bob share a separable state.

In the remaining cases Alice and Bob share a mixed entangled state. In order to distill entanglement from it, they can proceed in two ways. Firstly, if Alice and Bob share a large number of blocks of qubits in identical post measurement states, then they can apply the one-way hashing protocol \cite{Devetak1} and distill entanglement at the rates
\begin{eqnarray}
I_c(A>B)=S(B)-S(AB),
\end{eqnarray}
if Alice classicaly communicates to Bob or
\begin{eqnarray}
I_c(B>A)=S(A)-S(AB),
\end{eqnarray}
if Bob classically communicates to Alice. $S(A)$ and $S(B)$ are von Neumann entropies of Alice's and Bob's subsystems, respectively and $S(AB)$ is von Neumann entropy of the whole system. They are given by the following formulae:
\begin{eqnarray}
S(A)=\log_2{n \choose a},
\end{eqnarray}
because the state of Alice's subsystem is an equal mixture of all sequences of length $n$ with the Hamming weight equal to $a$;
\begin{eqnarray}
S(B)=\log_2{n \choose b},
\end{eqnarray}
because the state of Bob's subsystem is an equal mixture of all sequences of length $n$ with the Hamming weight equal to $b$;
\begin{eqnarray}
S(AB)=\log_2{n \choose a+b}
\end{eqnarray}
because the state of the whole system is an equal mixture of ${n \choose a+b}$ pure orthogonal states.
Hence, the optimal rate of the one-way hashing protocol is
\begin{eqnarray}
& I_{c}(n,a,b)=\nonumber\\
& = \log_2{[\max{\{{n \choose a},{n \choose b}\}]}}-\log_2{{n \choose n-a-b}}.
\end{eqnarray}

Secondly, Alice and Bob can divide the pairs of qubits into two blocks of equal length (the first block consists of the first $n/2$ pairs and the second block consists of the last $n/2$ pairs) and repeat the measurements on each block separately. If Alice obtains $P_{a'}^{n/2} \otimes P_{a''}^{n/2}$ as the result of her measurement and Bob obtains $P_{b'}^{n/2} \otimes P_{b''}^{n/2} $ as the result of his measurement, then the post measurement state is
\begin{eqnarray}
\rho(n/2,a',b')\otimes\rho(n/2,a'',b''),
\label{eq:4}
\end{eqnarray}
where $a'+a''=a$ and $b'+b''=b$. In derivation we used the identity 
\begin{eqnarray}
(P_{x'}^{n/2}\otimes P_{x''}^{n/2}) P_{x}^{n}=(P_{x'}^{n/2}\otimes P_{x''}^{n/2}) \delta_{x'+x'',x}.
\end{eqnarray}
The probability of this event is
\begin{eqnarray}
p(a',b';a'',b''|n,a,b)= \frac{p(n/2,a',b')p(n/2,a'',b'')}{p(n,a,b)}.
\end{eqnarray}
If $a'+b'=n/2$ ($a''+b''=n/2$), then the first (second) block of pairs is in the maximal entangled state of the rank $r(n/2,a')$ ($r(n/2,a'')$). If $a'+b'\neq n/2$ ($a''+b''\neq n/2$), then Alice and Bob can choose if they want to apply the one-way hashing protocol or to divide the pairs in the first (second) block into two blocks and repeat the measurement on each block separately. For different choices Alice and Bob obtain different rates. The rates achievable with the optimal choices are given by the following recursive formula:
\begin{widetext}
\begin{eqnarray}
& R(n, a, b)=\text{max}\{I_c(n,a,b),\nonumber\\
& \sum_{a'=a'_{\text{min}}}^{a'_{\text{max}}}\sum_{b'=b'_{\text{min}}}^{b'_{\text{max}}} p(a',b';a-a',b-b'|n,a,b)(R(n/2, a', b')+R(n/2,a-a',b-b'))\}
\end{eqnarray}
\end{widetext}
where the summation limits are
\begin{eqnarray}
& a'_{\text{min}}=\text{max}\{0,a-\frac{n}{2}\}\nonumber\\
& a'_{\text{max}}=\text{min}\{a,\frac{n}{2}\}\nonumber\\
& b'_{\text{min}}=\text{max}\{0,a+b-a'-\frac{n}{2}\}\nonumber\\
& b'_{\text{max}}=\text{min}\{b,\frac{n}{2}-a'\}
\end{eqnarray}

If we use only the bisection method (without the one-way hashing protocol), then we obtain the following expression for the rate of such a protocol for state $\rho$:
\begin{equation}
R(\rho)=\sum_{l=1}^{m}
p^{(2^l)} \big[Y(2^l)-Y(2^{l-1})\big],
\label{eq: yield}
\end{equation}
where $n=2^m$ and
\begin{equation}
Y(x)\equiv\frac{1}{x2^x}\sum_{k=0}^{x}\binom{x}{k}\log_2\binom{x}{k}.
\end{equation}

\begin{figure}[htb]
\includegraphics[angle=0, height=5cm, width=8cm]{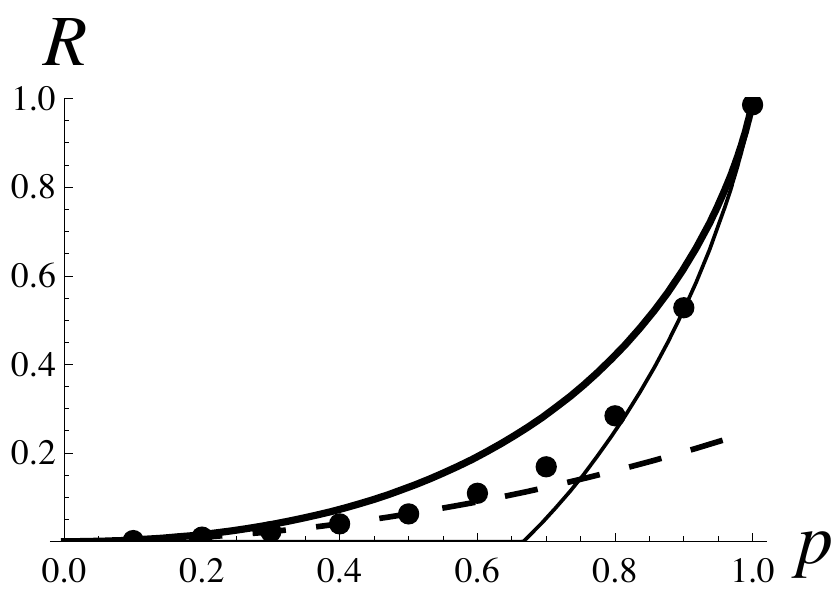}
\caption{Relation between coherent information (thin solid line), relative entropy of entanglement (thick solid line), the rate of the protocol of Bennett \emph{et al.} (dashed line) and our protocol for $n=64$ (dots) for different values of the parameter $p$.}
\label{fig:figure1}
\end{figure}

In Fig.~\ref{fig:figure1}  we present the rates of the protocol based on the bisection method and the one-way hashing protocol for different initial states $\rho$ of Eq.~\ref{eq:6} which depend on the parameter $p$. The first measurement was performed on a block of $64$ pairs of qubits. For comparison, we also present the rates of the protocol from \cite{Bennett3} and the one-way hashing protocol as well as an upper bound for distillable entanglement given by the relative entropy of entanglement. One can see that our protocol performs better than the other two protocols.

In Table~\ref{tab:table1} we present the rates of the protocol for different sizes of blocks of pairs of qubits on which the first measurement was performed. For comparison, we present also the rates of the protocol based only on the bisection method (without the one-way hashing protocol). One can see that the one-way hashing protocol causes an increase in the rate.

\begin{table}[htb]
\caption{Results for $p=\frac{2}{3}$. $n$ -- size of the block of pairs of qubits on which the first measurement is performed. $R$ -- the rate of the protocol based on bisection method and one-way hashing protocol. $R'$ -- the rate of the protocol based only on bisection method.}
\label{tab:table1}
\begin{center}
\begin{tabular}{|c|c|c|}\hline
 $n$& $R$ & $R'$\\ \hline
$2$& $0.111111$ & $0.111111$\\ \hline
$4$& $0.158981$ & $0.158981$\\ \hline
$8$& $0.173419$ & $0.16638$\\ \hline
$16$& $0.175076$ & $0.166574$\\ \hline
$32$& $0.175129$ & $0.166575$\\ \hline
$64$& $0.175129$ & $0.166575$\\ \hline
\end{tabular}
\end{center}
\end{table} 

Let us also point out that our protocol applies to a mixture of an arbitrary pure entangled state and a pure  product state orthogonal to it, i.e.,

\begin{equation}
\rho=p|\psi\rangle\langle\psi|+(1-p)|00\rangle\langle 00|
\label{eq:5}
\end{equation}
where 
\begin{equation}
|\psi\rangle=\alpha|10\rangle+\beta|01\rangle
\end{equation}

One can see it by noting that the first measurement projects $n$ copies of the initial state on a state given in Eq.~\ref{eq:1} with probability
\begin{eqnarray}
& P'(n,a,b)=p^{a+b}(1-p)^{n-a-b}{n\choose a+b} \times \nonumber\\
& \times {a+b\choose a}|\alpha|^{2a}|\beta|^{2b}.
\end{eqnarray}

Because the states given in Eq.~\ref{eq:5} can be obtained by sending half of the state
\begin{equation}
|\psi'\rangle=\alpha'|10\rangle+\beta'|01\rangle)
\end{equation}
through the amplitude damping channel $N_{AD}$, given by Kraus operators
\begin{eqnarray}
& E_{0}=|0\rangle \langle0|+\sqrt{1-\gamma^2}|1\rangle \langle 1|\nonumber\\
& E_{1}=\gamma |0\rangle \langle1|
\end{eqnarray}
we can obtain the lower bound on quantum capacity of the amplitude damping channel assisted by two-way classical communication $Q_2$. In such a case, the best lower bound on $Q_2$ is given by the following expression
\begin{eqnarray}
R_{\max} = \max_{\psi'}R(N_{AD}(\psi')).
\end{eqnarray}

Our protocol requires collective measurements on large blocks of pairs of qubits, while the protocol presented in \cite{Bennett3} requires only measurements on two pairs of qubits. We will show how one can improve the two-copy protocol. Let Alice and Bob perform von Neumann measurements given by projectors
\begin{eqnarray}
& P_0=|00\rangle\langle00|+|11\rangle\langle11|\nonumber\\
& P_1=|01\rangle\langle01|+|10\rangle\langle10|
\end{eqnarray}
on the state $\rho^{\otimes 2}$
If both obtain $1$ as the results of their measurements, then the post measurement state is equivalent to the maximal entangled state $|\psi^+\rangle$. The probabilty of this event is equal to $\frac{p^2}{2}$. If both obtain $0$ as the results of their measurements, then the post measurement state is equivalent to
\begin{eqnarray}
\rho'=p'|\psi^{+}\rangle\langle\psi^{+}|+(1-p')|00\rangle\langle 00|
\end{eqnarray}
where $p'=\frac{p^2}{p^2 + 2 (1-p)^2}$. The probability of this event is equal to $\frac{p^2}{2} + (1-p)^2$. Two such states can be used in another measurement and hence there is a chance to obtain from them the maximal entangled state. In particular, if $p=2/3$ then $p'=2/3$ and the rate is equal to $2/15$ for improved protocol instead of $1/9$ for the original protocol.

In conclusion, we presented the entanglement purification protocol for a mixture of a pure entangled state and a pure product state, orthogonal to each other. To our knowledge uur protocol performs better than any other previously known protocol. We also discussed how one can obtain the lower bound on quantum capacity of the amplitude damping channel assisted by two-way classical communication.

\begin{acknowledgments}
Fruitful discussion with M. Horodecki is acknowledged. One of the authors (A.G.) was partially supported by Ministry of Science and Higher
Education Grant No. N N206 2701 33 and by the European Commission through the Integrated Project FET/QIPC SCALA.

\end{acknowledgments}

\end{document}